\documentclass[fleqn,twoside]{article}
\usepackage{espcrc2}
\usepackage{epsfig}
\usepackage{amssymb}
\usepackage[figuresright]{rotating}

\newcommand{\AmS}{{\protect\the\textfont2
  A\kern-.1667em\lower.5ex\hbox{M}\kern-.125emS}}

\newcommand{\ksea}{\mbox{$\kappa^{\rm sea}$}}
\newcommand{\kval}{\mbox{$\kappa^{\rm val}$}}

\newcommand{\err}[2]{${\scriptstyle {}^{+{#1}}_{-{#2}}}$}

\title{Searching for dynamical fermion effects in UKQCD simulations}

\author{Chris Allton\address{Department of Physics, University of Wales Swansea, U.K.}\\
\hspace{10mm}for the {\it UKQCD Collaboration}}

\begin{document}

\begin{abstract}
We present recent results from the UKQCD collaboration's dynamical QCD
simulations. This data has fixed lattice spacing but varying dynamical
quark mass. We concentrate on searching for an unquenching signal in
the mesonic mass spectrum where we do not find a significant effect at
the quark masses considered.
\end{abstract}

\maketitle

%}}}

%{{{ Introduction

\section{Introduction}

Computing resources available for simulations of lattice QCD
are now powerful enough to investigate unquenching effects.
UKQCD has embarked on a programme of studying these effects in the
light hadron spectrum, static quark potential, glueball spectrum
and topological sectors. The purpose of this paper is to investigate
unquenching effects in the first of these areas.

It is well known that the lattice cut-off is a function of {\em both}
the gauge coupling, $\beta$, and dynamical quark mass, $m^{sea}$ (see
e.g. \cite{csw176}). For this reason, the philosophy we have chosen is
to simulate at points along the ``matched'' trajectory in the
$(\beta,m^{sea})$ plane i.e. defined by{\em fixed} lattice spacing,
$a$. This then disentangles lattice spacing artefacts with unquenching
effects. Any variation of a physical quantity along this trajectory
can sensibly be attributed to unquenching effects rather than lattice
systematics. This work has been published in full in \cite{csw202}.

%}}} 

%{{{        table

\begin{table*}[tb]
\caption{Lattice parameters together with measurements of $a$ and quark masses.}
\label{tb:params}
\begin{center}

\begin{tabular}{lllllll}
$\beta$ & $\ksea$ & $c_{SW}$ & \#conf. &  
\multicolumn{1}{c}{ $a_{r_0}$ [Fm]} &
\multicolumn{1}{c}{ $a_{J}$ [Fm]} &
\multicolumn{1}{c}{ $M_{PS}^{unitary}/M_V^{unitary}$}
 \\
\hline
5.20& 0.1355 &2.0171 & 208 & 0.0972(8)\err{7}{0}   & 0.110\err{ 4}{ 3}& 0.578\err{13}{19}\\
\hline
5.20& 0.1350 &2.0171 & 150 & 0.1031(09)\err{20}{1} & 0.115\err{ 3}{ 3}& 0.700\err{12}{10}\\
5.26& 0.1345 &1.9497 & 101 & 0.1041(12)\err{11}{10}& 0.118\err{ 2}{ 2}& 0.783\err{ 5}{ 5}\\
5.29& 0.1340 &1.9192 & 101 & 0.1018(10)\err{20}{7} & 0.116\err{ 3}{ 4}& 0.835\err{ 7}{ 7}\\
\hline 
5.93& 0      &1.82   & 623 & 0.1040(03)\err{4}{0}  & 0.1186\err{17}{15}& 1   \\
\end{tabular}
\end{center}
\end{table*}

%}}}

%{{{ Simulation Details

\section{Simulation Details}

A non-perturbatively improved clover action was used with
lattice parameters displayed in Table~\ref{tb:params}
with a volume of $16^3 \times 32$.
The last four rows contain the parameters for
the matched ensembles, while the top simulation
explores a lighter quark mass.
Further details of the simulation appear in \cite{csw202,derek}.

%}}}

%{{{ Mesonic Sector

\section{Mesonic Sector}

We begin outlining our results by indicating our quark masses
via $M_{PS}^{unitary}/M_V^{unitary}$, where
$M_{V(PS)}$ is the pseudoscalar (vector) meson mass and
the superscript
``unitary'' refers to the parameters $m^{sea} \equiv m^{val}$
in Table~\ref{tb:params}.
As can be seen our quark masses are only modestly light
(c.f. $M_{PS}^{unitary}/M_V^{unitary} \approx 0.18$ in nature).

The results for the hyperfine splitting are shown in
Fig.~1 where the experimental points are plotted
assuming $r_0 = (0.49 \pm 0) fm$.
Note that there is a tendency for the lattice data to flatten towards
the experimental points as the dynamical quark mass decreases within
the matched ensembles.
However the ``unmatched'' simulation ($\ksea=0.1355$) shows an
increased negative slope, presumably due to finite-volume effects.

%{{{        fig:hyperfine_r0.eps

\begin{figure}[htp]
\begin{center}
\epsfig{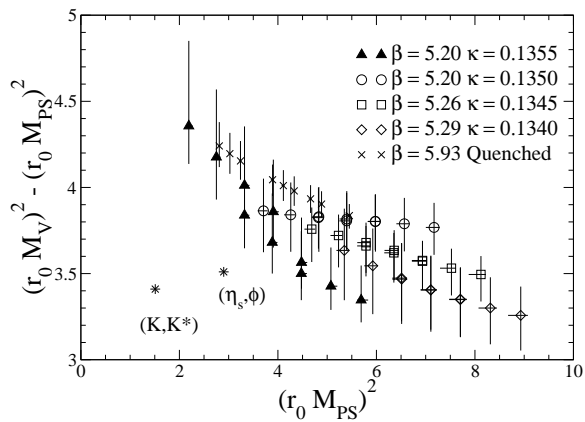}
\end{center}
\label{fig:hyperfine}
\vspace{-10mm}
\caption{The hyperfine splitting.}
\vspace{-3mm}
\end{figure}

%}}}

The $J$ parameter, defined as,
\begin{equation}
J = M_V \frac{dM_V}{dM_{PS}^2} \bigg|_{K,K^\ast},
\label{eq:J}
\end{equation}
was calculated using 3 approaches:
(i) ``Pseudo-Quenched'' i.e. where $\ksea$ is held fixed
and the derivative in eq.(\ref{eq:J}) is w.r.t. variations
in $\kval$;
(ii) ``Unitary Trajectory'' i.e. where $\ksea \equiv \kval$
and the derivative in eq.(\ref{eq:J}) is w.r.t. variations
in both $\kval$ and $\ksea$ combined; and
(iii) ``Chiral Extrapolation of Pseudo-Quenched''
i.e. taking the $J$ values from Approach (i) and performing the
chiral extrapolation $m^{sea} \rightarrow 0$.
The results of all three methods are shown in Fig.~2,
together with the experimental point.
There is evidence that the lattice value for $J$ approaches
the experimental point as the sea quark mass decreases
(see approaches 1 and 3). Note also that the ``unmatched''
simulation ($\ksea=0.1355$) has a {\em smaller} $J$ value.
This fact is related to the comment in the previous paragraph.

%{{{        fig:J.eps

\begin{figure}[htp]
\begin{center}
\epsfig{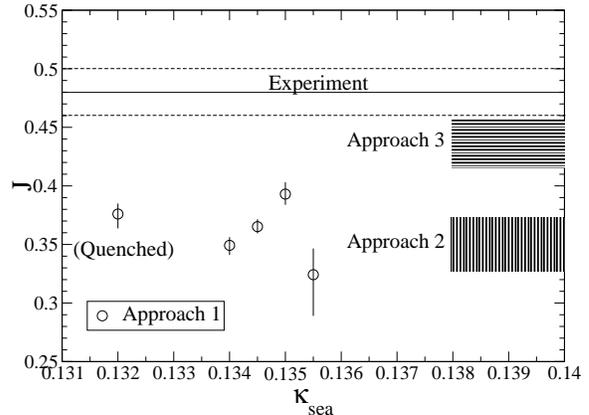}
\end{center}
\label{fig:J}
\vspace{-10mm}
\caption{The $J$ parameter from the three methods
described in the text.}
\vspace{-3mm}
\end{figure}

%}}}

We use two methods to extract the lattice spacing.
The first is via the Sommer scale $r_0$, and the second is
directly from the meson spectrum as outlined in \cite{leonardo}.
The values obtained from both methods are displayed in
Table~\ref{tb:params}. Note that $a_J$ is $\approx$ 10\% larger
than $a_{r_0}$. This could be due to the experimental estimate of
$r_0 = 0.49$ fm being $\approx$ 10\% too small.

%}}}

%{{{ Unquenching Effects in Meson Spectrum

\section{Unquenching Effects in Meson Spectrum}

In order to investigate unquenching effects in the meson
spectrum, we define the quantity
\[%\begin{equation}
\delta_{i,j}(\beta,\ksea) = 1 - \frac{a_i(\beta,\ksea)}
                                     {a_j(\beta,\ksea)},
\]%\end{equation}
where $a_i$ is the scale determined from the physical
quantity $M_i$.
When $\delta_{i,j} = 0$ then the lattice prediction
of $M_i$ with scale taken from $M_j$ agrees with experiment.
Thus $\delta$ is a good parameter to study unquenching effects.
We expect that $\delta_{i,j}(\beta,m^{sea}) = {\cal O}(a^2)$
since we are using a non-perturbatively improved clover action.

In Fig.~3 we plot $\delta_{i,j}$ for the matched datasets
where $j=r_0$, and $i$ is the scale determined from the string tension,
$\sigma$, and the two mass pairs $(\rho,\pi)$ and $(K,K^\ast)$
following \cite{leonardo}.
The $x-$axis in this plot is
$(aM_{PS}^{unitary})^{-2} \sim 1/m^{sea}$, so the quenched data point
lies on the $y-$axis.

%{{{        fig:delta

\begin{figure}[htp]
\begin{center}
\epsfig{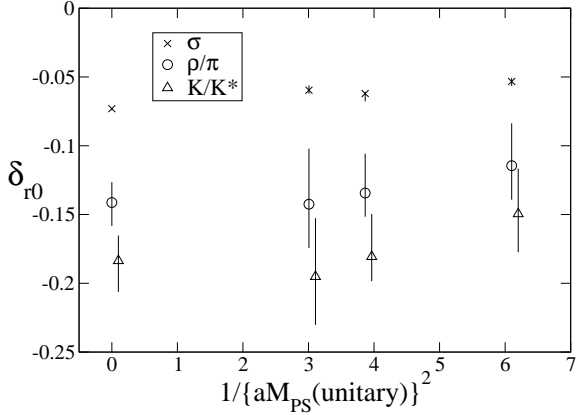}
\end{center}
\label{fig:delta}
\vspace{-10mm}
\caption{$\delta_i$ as described in the text.}
\vspace{-3mm}
\end{figure}

%}}}

We see disappointingly that, in the case of the scale determinations
from the meson spectrum, there is no significant tendency of $\delta$
towards zero as $m^{sea} \rightarrow 0$, i.e. the quantity $\delta$
does not give us an indication of unquenching effects. However, in the
case of $\sigma$, there is a statistically significant variation of
$\delta \rightarrow 0$ as $m^{sea} \rightarrow 0$. This implies that
we do see unquenching effects in the static quark potential.

%}}}

%{{{ Chiral Extrapolations

\section{Chiral Extrapolations}

In \cite{csw202} we used 3 approaches to perform the chiral
extrapolations of hadron masses: 
``Pseudo-Quenched'';
``Unitary Trajectory'';
and a ``Combined Chiral Fit''.
In this publication we will refer only to the last approach.
We take
\[
\hat{M}(\ksea;\kval)
\]
\[
\;\;\;\;\;\;\;\;\;\;\;\;= A(\ksea) + B(\ksea) \hat{M}_{PS}(\ksea;\kval)^2
\]
\[
\;\;\;\;\;\;\;\;\;\;\;\;= A_0 + A_1 \hat{M}_{PS}(\ksea;\ksea)^{-2}
\]
\[
+ \left[ B_0 + B_1 \hat{M}_{PS}(\ksea;\ksea)^{-2} \right]
\hat{M}_{PS}(\ksea;\kval)^2,
\]
using the nomenclature $\hat{M} \equiv aM$ and
the first argument of $M(\ksea;\kval)$ refers
to the sea quark and the second to the valence quark.

The results of these extrapolations are shown in
Table~\ref{tb:chiral}. We stress that this functional form for the
extrapolation is not motivated by theory, but is used as a numerical
analysis technique in order to test for evidence of
unquenching effects. As can be seen from Table~\ref{tb:chiral}, the
parameters $A_1$ and $B_1$ are compatible with zero (to $2\sigma$)
and therefore we conclude that there is no evidence of unquenching
effects.

%{{{    table combined

\begin{table}[*htbp]
\begin{center}
 \begin{tabular}{lcccc}
hadron  & $A_0$ & $A_1$ & $B_0$ & $B_1$ \\
\hline
Vector       & .492\err{10}{ 9} & -0.004\err{ 2}{ 3} & 0.61\err{ 4}{ 4} &  .015\err{ 9}{ 7} \\
Nucleon      & .663\err{13}{15} &  0.006\err{ 3}{ 4} & 1.23\err{ 6}{ 6} & -.001\err{1}{1} \\
Delta        & .84\err{ 2}{ 2}  & -0.002\err{ 5}{ 5} & 0.91\err{ 8}{ 9} &  .02\err{ 2}{ 2} \\
\end{tabular}
\end{center}
\caption{Fit parameters from the Chiral Extrapolations}
\label{tb:chiral}
\end{table}

%}}}

%}}}

%{{{ Conclusions

\section{Conclusions}

This paper attempts to uncover unquenching effects in the dynamical
lattice QCD simulations at a fixed (matched) lattice spacing (and
volume) and various dynamical quark masses. This approach allows
a more controlled study of unquenching effects without the possible
entanglement of lattice and unquenching systematics.

However, we see no significant sign of unquenching effects in the
meson spectrum. This is presumably since our dynamical quarks are
relatively massive, and so the meson spectrum is dominated by the
static quark potential. This potential is, by definition, matched
amongst our ensembles at the hadronic length scale $r_0$, and so
any variation of the meson spectrum within our matched ensemble
must surely be a ``higher'' order unquenching effect which is beyond
our present statistics.

We have however shown that unquenching effects exist in the static
quark potential, and other work (using the same ensembles) has shown
interesting unquenching effects in the glueball and topological sector
\cite{csw202}.

%}}}

%{{{ Bibliography

%}}} 

\end{document}